**Coulomb Blockade in a Silicon/Silicon-Germanium Two-Dimensional Electron Gas Quantum Dot**

L.J. Klein, K.A. Slinker, J.L. Truitt, S. Goswami, K.L.M. Lewis, S.N. Coppersmith, D.W. van der Weide, Mark Friesen, R.H. Blick, D.E. Savage, M.G. Lagally, Charles Tahan, Robert Joynt, M.A. Eriksson[1]
*University of Wisconsin Madison, Madison, WI 53706*

J.O. Chu, J.A. Ott, P.M. Mooney
*IBM Research Division, T. J. Watson Research Center, NY 10598*

**Abstract:**
We report the fabrication and electrical characterization of a single electron transistor in a modulation doped silicon/silicon-germanium heterostructure. The quantum dot is fabricated by electron beam lithography and subsequent reactive ion etching. The dot potential and electron density are modified by laterally defined side gates in the plane of the dot. Low temperature measurements show Coulomb blockade with a single electron charging energy of 3.2 meV.

Silicon-germanium modulation doped field-effect transistors (MODFETs) are potentially attractive devices for high-speed, low noise communications applications, where low cost and compatibility with CMOS logic are desirable.[1] Because the silicon quantum well containing the electrons is strained by up to 2%, the electron mobility of these structures is as much as a factor of five larger than that of unstrained silicon field-effect transistors (FET) at room temperature, offering the prospect of high speed operation. At low temperatures, electron mobilities as high as $5.2 \times 10^5$ cm$^2$/Vs have been reported,[2,3] raising the possibility of lithographically patterned quantum devices.

Development of quantum devices in silicon MODFETs is of particular interest, because silicon is unique among the elemental and binary semiconductors in that it has an abundant nuclear isotope of spin zero. Silicon also has very small spin orbit coupling. Together, these two features provide only weak channels for electron spin relaxation; the electron spin dephasing time $T_2$ for phosphorus-bound donors has been measured to be as long as 3 ms at 7 K.[4] Kane has pointed out the advantages of nuclear spins in silicon for quantum computation,[5] and his scheme has been extended to electrons in SiGe heterostructures.[6] Following Loss and DiVincenzo,[7] specific schemes have been proposed for spin-based quantum computation in silicon-germanium electron quantum dots.[8,9]

Here we demonstrate a quantum dot fabricated in a layered silicon/silicon-germanium (Si/SiGe) heterostructure that includes a strained Si quantum well containing a two-dimensional electron gas (2DEG). Even with recent advances in the growth of high mobility SiGe modulation-doped heterostructures, producing lithographically defined n-type quantum dots with periodic Coulomb blockade has been challenging. Particularly difficult is the fabrication of highly isolated Schottky top gates.[10,11] Due to the lattice mismatch between layers of different Ge fraction, misfit dislocations must be present to relieve the strain in the SiGe buffer layer. Misfit dislocations terminate in threading arms running up to the heterostructure surface, and these threading arms may play a role in forming a conductive path between top Schottky contacts and the 2DEG below. We have avoided this problem by fabricating a dot with highly isolated *side* gates formed from the 2DEG itself.

---
[1] Electronic mail: maeriksson@wisc.edu

The Si/SiGe heterostructure used here was grown by ultra-high vacuum chemical vapor deposition (UHVCVD).[2] The 2DEG sits near the top of 80 Å of strained Si grown on a strain-relaxed $Si_{0.7}Ge_{0.3}$ buffer layer, as shown in Fig. 1(a). The 2DEG is separated from the donors by a 140 Å $Si_{0.7}Ge_{0.3}$ spacer layer, and the phosphorus donors lie in a 140 Å $Si_{0.7}Ge_{0.3}$ layer capped with 35 Å of Si at the surface. The electron density in the 2DEG is $4 \times 10^{11}$ cm$^{-2}$ and the mobility is 40,000 cm$^2$/Vs at 2K. Ohmic contacts to the 2DEG are formed by Au/Sb metal evaporation and sintering at 400 °C for 10 min.

Quantum dots are fabricated by electron beam lithography and subsequent $CF_4$ reactive-ion etching. Fig. 1(b) shows an atomic force microscope image of the completed structure. Control of the dot electron population and the lead resistances is achieved with three separately tunable gates. Each gate is fabricated from the same 2DEG from which the quantum dot is created. Such in-plane coupling of one 2DEG to another has been used to monitor the electron population in gallium-arsenide quantum dots.[12] Here we invert this idea and use the 2DEG-2DEG coupling to control the dot dimensions. The data presented in this letter were acquired at 1.8 K and 250 mK, during different cool-downs of the sample. The general electronic properties of the dot were similar on each cool-down, although the detailed Coulomb blockade peak positions differed.

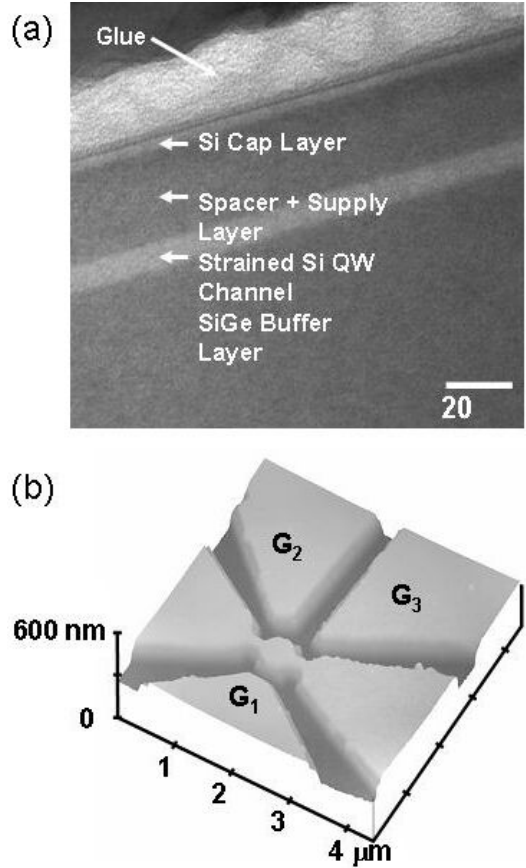

**Fig. 1.** (a) Transmission electron microscope image of the SiGe heterostructure used in this work. The 2DEG sits near the top of the silicon quantum well. (b) Atomic force microscope image of the fabricated dot with three etch-defined electrostatic side gates.

A current-voltage (I-$V_{ds}$) measurement of the dot at zero gate bias taken at 250 mK is shown in Figure 2(a). The leads of the dot are intrinsically pinched off in the tunneling regime, due to surface depletion, such that the Coulomb blockade region is evident at zero gate voltage for $|V_{ds}| < 4$ mV. The dot remains in the tunneling regime up to + 5V applied to the side gates. Conductance oscillations with varying gate voltage are observed for each of the three gates. Typical results are shown in Figure 2(b-d) at 1.8 K. A standard lock-in technique is used for conductance measurements with a 100 µV ac voltage applied between source and drain at 19 Hz. The periodic spacing of the peaks for the larger gate is $\Delta V_{g1}$=125 mV . The two smaller gates show correspondingly larger periods ($\Delta V_{g2}$=155 mV, $\Delta V_{g3}$=226 mV). The spacing of the oscillations implies gate-dot capacitances of $C_{g2}$=1.03 aF and $C_{g3}$=0.71 aF for the smaller, more distant gates, and $C_{g1}$=1.28 aF for the larger, closer gate. These capacitances are smaller than would be expected for top metal gates because the side gates are in the plane of the dot and are spaced farther away due to the etching necessary to define them. Furthermore, in these configurations some of the electric field lines between the side gates and the dot travel through the air gap (lower dielectric constant) rather than through the SiGe heterostructure. The current flowing through the quantum dot vanishes for some, but not all of the minima between Coulomb blockade peaks. In addition, the minima seem to come in small bunches, with each bunch displaying either deep or shallow minima (Fig. 2). This multiple periodicity may be an indication of a small disorder-induced dot near the primary etch-defined quantum dot. This disorder-induced dot may be a source of parallel conduction under certain ranges of the gate voltages, leading to the shallow minima between some sets of Coulomb blockade peaks. Fig. 3 presents Coulomb blockade oscillations through the dot at various temperatures. A broadening and a general increase in the background are apparent at higher temperature, as expected for Coulomb blockade.

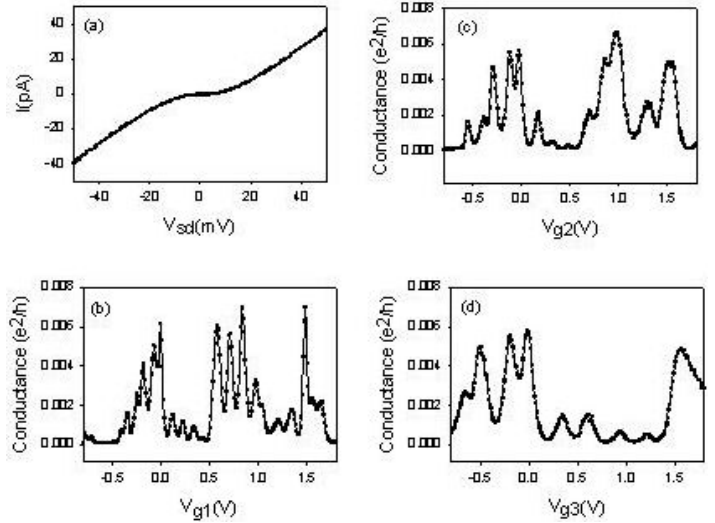

**Fig. 2. (a)** I-V curve through the dot at zero gate bias showing zero conductance up to 4 mV source-drain voltage. **(b)-(d)** Coulomb blockade oscillations through the quantum dot as the voltage is varied on each of the three side gates ($G_1$, $G_2$, $G_3$) respectively.

Figure 4 is a two-dimensional plot showing dI/dV measurements with varying gate and drain-source voltages (Coulomb diamonds) for the device. The data were acquired at 250 mK with an 80 µV ac voltage applied between drain and source at 200 Hz. The charging energy to overcome the Coulomb blockade and add an electron to the dot can be estimated from this plot using[13]

$$E_C = e\frac{C_g}{C}\Delta V_g = \frac{1}{2}e\left(\frac{dV_{ds}}{dV_g}\right)\Delta V_g$$

where $dV_{ds}/dV_g$ is the slope of the diamonds and $\Delta V_g$ is the spacing of the Coulomb oscillations obtained above (Fig. 2). For small gate voltages (the left side of the plot) the charging energy obtained is 3.2 meV. On the right side of the Coulomb diamond plot the diamonds do not close completely. This can occur in the presence of disorder in the 2DEG, in which case the conduction can be impeded by charging of trap states or smaller dots – effectively creating multiple quantum dots in this gate voltage range. Also apparent in the diamond plot are a few switching events around 0.6V in gate voltage due to trapped charge rearrangement. However, the periodicity of the conduction oscillations remains apparent in the stability diagram, indicating that the fabricated quantum dot is still dominating the transport.

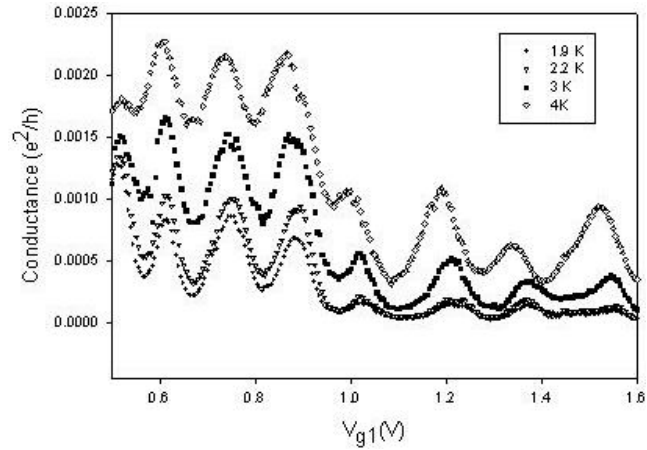

**Fig. 3.** Conductance oscillations through the quantum dot as a function of temperature as the voltage on gate 3 is varied.

The total capacitance of the dot as calculated from this charging energy is 50 aF. This corresponds to a disk of diameter 120 nm in an infinite dielectric. A better estimate can be made with Poisson simulations of the full device, using an adaptive finite element mesh, and treating 2DEG regions (dot, leads, and gates) with Dirichlet boundary conditions. We thus estimate an electronic dot diameter of $D = 233$ nm. This result compares favorably to independent measurements of ~200 nm surface depletion in quantum wires of variable width that were fabricated in a similar manner.[14] From the electronic dot diameter and the sheet density of electrons in the original 2DEG, we estimate that there are ~170 electrons in the dot under the operating conditions of Figures 2 and 3. Fabrication of smaller dots using the etch-defined gates described here will allow lower electron occupation of the dot. It is likely that achievement of individual electron quantum dots will require either etch-defined gates that are more closely coupled than those demonstrated here, or the use of metal top gates to confine the electrons laterally.

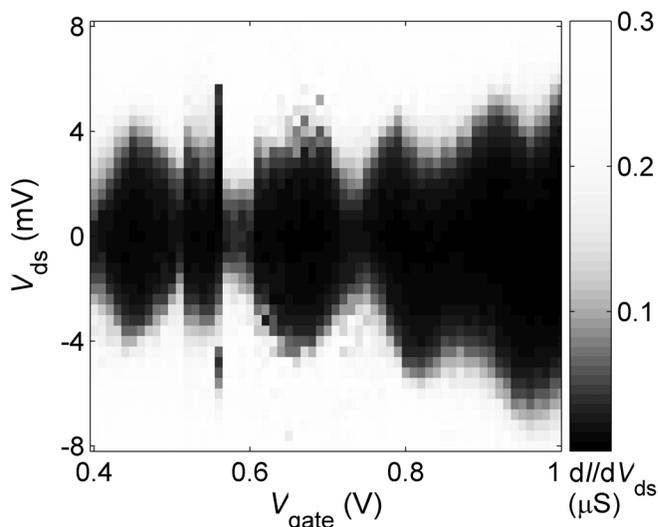

**Fig. 4.** Stability plot of the differential conductance through the dot as a function of the voltage on gate 1 and the source-drain voltage, at a temperature of 250 mK.

In conclusion, a single electron transistor in an n-type SiGe heterostructure was fabricated. The potential of the dot is modulated by side gates defined by etching and Coulomb blockade behavior is observed. Over a wide range in gate voltage (Fig. 2), single dot Coulomb blockade is observed. The dot is shown to be stable over moderate time periods with varying gate and drain-source voltages. In this work we have employed traditional low frequency lock-in techniques. A long term goal is the manipulation of silicon dots at much higher frequencies. Operation of the quantum dot at high frequencies requires either a wide bandwidth current preamplifier, possibly operated at low temperatures, or detection of charge motion in and out of the dot by a radio-frequency single electron transistor in the proximity of the fabricated quantum dot.[15,16]

We acknowledge K. Ismail and K.L. Saenger for contributions to the development of these high mobility heterostructures, and support from ARDA, ARO, and NSF.

**References:**
[1] S.J. Koester, K.L. Saenger, J.O. Chu, Q.C. Ouyang, J.A. Ott, M.J. Rooks, D.F. Canaperi, J.A. Tronello, C.V, Jhanes and S.E. Steen, Electronics Letters (in press).
[2] K. Ismail, M. Arafa, K. L. Saenger, J. O. Chu, and B. S. Meyerson, Appl. Phys. Lett. **66**, 1077 (1995).
[3] K. Ismail, M. Arafa, F. Stern, J. O. Chu, and B. S. Meyerson, Appl. Phys. Lett. **66**, 842 (1995).
[4] Tyryshkin, A.M., Lyon, S.A., Astashkin, A.V., Raitsimring, A.M., preprint cond-mat/0303006.
[5] B. E. Kane, Nature **393**, 133 (1998).
[6] Rutger Vrijen, Eli Yablonovitch, Kang Wang, Hong Wen Jiang, Alex Balandin, Vwani Roychowdhury, Tal Mor, and David DiVincenzo, Phys. Rev. A **62**, 12306 (2000).
[7] D. Loss and D. P. DiVincenzo, Phys. Rev. A **57**, 120 (1998).
[8] J. Levy, Phys. Rev. A **64**, 052306 (2001).

9   M. Friesen, P. Rugheimer, D. E. Savage, M. G. Lagally, D. W. van der Weide, R. Joynt, and M. A. Eriksson, Phys. Rev. B **67**, 121301 (2003).
10  S. Kanjanachuchai, J. M. Bonar, and H. Ahmed, Semiconductor Science and Technology **14**, 1065 (1999).
11  A. Notargiacomo, L. Di Gaspare, G. Scappucci, G. Mariottini, F. Evangelisti, E. Giovine, and R. Leoni, Appl. Phys. Lett. **83**, 302 (2003).
12  J. M. Elzerman, R. Hanson, J. S. Greidanus, L. H. W. van Beveren, S. De Franceschi, L. M. K. Vandersypen, S. Tarucha, and L. P. Kouwenhoven, Phys. Rev. B **67**, 161308 (2003).
13  L. P. Kouwenhoven, C. M. Marcus, P. L. McEuen, S. Tarucha, R. M. Westervelt, and N. S. Wingreen, in *Mesoscopic Electron Transport*, edited by L.L. Sohn, L.P. Kouwenhoven, and G. Schön (Kluwer, 1997), Vol. 345, p. 105.
14  K.L.M. Lewis, K. Slinker, L.J. Klein, and M.A. Eriksson unpublished.
15  R. J. Schoelkopf, P. Wahlgren, A. A. Kozhevnikov, P. Delsing, and D. E. Prober, Science **280**, 1238 (1998).
16  W. Lu, Z. Q. Ji, L. Pfeiffer, K. W. West, and A. J. Rimberg, Nature **423**, 422 (2003).